\documentclass[aps,prl,reprint,superscriptaddress,showpacs]{revtex4-1}
\usepackage{graphicx}

\begin{document}

\title{Giant Phonon Softening and Enhancement of Superconductivity by Phosphorus Doping of BaNi$_2$As$_2$}

\author{K. Kudo}
\email{kudo@science.okayama-u.ac.jp}
\author{M. Takasuga}
\affiliation{Department of Physics, Okayama University, Okayama 700-8530, Japan}
\affiliation{Transformative Research-Project on Iron Pnictides (TRIP), Japan Science and Technology Agency (JST),
5, Sanbancho, Chiyoda, Tokyo 102-0075, Japan}

\author{Y. Okamoto}
\author{Z. Hiroi}
\affiliation{Transformative Research-Project on Iron Pnictides (TRIP), Japan Science and Technology Agency (JST),
5, Sanbancho, Chiyoda, Tokyo 102-0075, Japan}
\affiliation{Institute for Solid State Physics, The University of Tokyo, Kashiwa 277-8581, Japan}

\author{M. Nohara}
\affiliation{Department of Physics, Okayama University, Okayama 700-8530, Japan}
\affiliation{Transformative Research-Project on Iron Pnictides (TRIP), Japan Science and Technology Agency (JST),
5, Sanbancho, Chiyoda, Tokyo 102-0075, Japan}

\date{April 19, 2012}

\begin{abstract}
The effects of phosphorus doping on the structural and superconducting phase transitions of BaNi$_2$(As$_{1-x}$P$_x$)$_2$ were studied. The specific heat, resistivity, and magnetic susceptibility were measured. The results revealed an abrupt increase in the superconducting transition temperature ($T_c$) from 0.6 K in the triclinic phase (space group P$\bar{1}$) with less phosphorus (0 $\leq$ $x$ $\leq$ 0.067) to 3.3 K in the tetragonal phase (space group I4/mmm) with more phosphorus ($x$ $\geq$ 0.067). Our data analysis suggests that a doping-induced softening related to an in-plane Ni and As(P) phonon mode is responsible for the enhanced superconductivity in the tetragonal phase.
\end{abstract}

\pacs{74.70.Xa, 74.25.Kc, 74.25.Dw, 74.25.-q, 74.25.Bt}

\maketitle

Superconductivity at high transition temperatures ($T_c$) is often associated with structural instabilities that are characterized by phonon softening and subsequent structural phase transitions. Enhancement in $T_c$ at a structural phase boundary has been observed in various systems including copper oxides \cite{PhysRevLett.88.167002,Nature04704}, iron pnictides \cite{JPSJ.80.073702,JPSJ.81.024604}, A15 compounds \cite{RevModPhys.47.637}, graphite intercalated compounds \cite{PhysRevB.74.024505,PhysRevB.74.214513,PhysRevLett.98.067002,PhysRevB.78.064506},
and elements such as Li and Te under high pressure \cite{PhysRevLett.96.047003,PhysRevLett.44.1623,Akahama1992803,PhysRevB.65.064504,PhysRevLett.77.1151,PhysRevB.63.224107,0953-8984-19-12-125206}.
For example, the graphite intercalated compound CaC$_6$ exhibits a gradual increase in $T_c$ from 11.5 to 15.1 K under high pressure followed by a sudden decrease down to $\simeq$ 5 K at 8$-$10 GPa \cite{PhysRevLett.98.067002}.
The temperature-dependent resistivity, together with first-principles calculations, suggest that the softening of in-plane Ca vibrations leads to the enhanced $T_c$ and the subsequent phase transition in CaC$_6$ under pressure \cite{PhysRevLett.98.067002,PhysRevB.74.214513,PhysRevB.78.064506}.
Tellurium (Te) exhibits a $T_c$ jump from 2.5 to 7.4 K at 32$-$35 GPa, which is related to the transition from the rhombohedral $\beta$-Po structure to the bcc phase \cite{PhysRevLett.77.1151}.
First-principles calculations suggest that the remarkable increase in $T_c$ can be attributed to a phonon softening of the transverse mode in the bcc phase under high pressure \cite{PhysRevLett.77.1151,PhysRevB.63.224107,0953-8984-19-12-125206}.
Thus, the tuning of soft phonons appears to be important for maximizing $T_c$. However, the chemical tuning of soft phonons, which is essential for practical applications, has not yet been fully investigated.

BaNi$_2$As$_2$ crystallizes in a tetragonal ThCr$_2$Si$_2$-type structure (space group I4/mmm) \cite{0953-8984-20-34-342203,PhysRevB.79.094508,PhysRevLett.102.147004}.
This system exhibits a structural phase transition from a tetragonal to a triclinic phase (space group P$\bar{1}$) at approximately 130 K, below which alternate Ni--Ni bonds  (with distances of $\simeq$ 2.8 {\AA} and $\simeq$ 3.1 {\AA}) are formed in the Ni square lattice \cite{PhysRevB.79.094508}. In this triclinic phase, superconductivity, considered to be of the conventional BCS type, emerges at 0.7 K  \cite{0953-8984-20-34-342203,PhysRevLett.102.147004,PhysRevB.78.132511,PhysRevB.79.054510}.
Therefore, this system can be considered a non-magnetic analogue of iron-based superconductors such as BaFe$_2$As$_2$.
In this Letter, we report that $T_c$ abruptly increases from 0.7 to 3.3 K upon the phosphorus doping of BaNi$_2$As$_2$.
The enhanced superconductivity is accompanied by the triclinic-to-tetragonal phase transition that is induced by phosphorus doping at $x$ = 0.067 in BaNi$_2$(As$_{1-x}$P$_x$)$_2$. The specific heat and electrical resistivity suggest that doping-induced phonon softening, related to the in-plane Ni and As vibrations, is responsible for the enhanced superconductivity, demonstrating that the chemical tuning of soft phonons is an effective means to optimize superconductivity.
%

Single crystals of BaNi$_2$(As$_{1-x}$P$_x$)$_2$ were grown using a self-flux method. A mixture with a ratio of Ba:NiAs:Ni:P = 1:$4(1-x)$:$4x$:$4x$ was placed in an alumina crucible, sealed in an evacuated quartz tube,
heated at 700 $^\circ$C for 3 h, and cooled from 1150 to 1000 $^\circ$C at a rate of 2 $^\circ$C/h, followed by furnace cooling. Single crystals with a typical dimension of 1 $\times$ 1 $\times$ 0.1 mm$^3$ were mechanically isolated from the flux. The results of powder X-ray diffraction, performed using a Rigaku RINT-TTR III X-ray diffractometer with Cu K$_\alpha$ radiation, showed that all specimens are in a single phase. Energy dispersive X-ray spectrometry (EDS) was used to determine the phosphorus content $x$.
Samples with $x$ greater than 0.13 could hardly be obtained by either flux growth or the solid-state reaction, indicating that the solubility limit of P for As is at $x$ = 0.13. The samples were treated in a glove box filled with dried Ar gas because they degraded in air within a few days.
The magnetization $M$ was measured using a Quantum Design MPMS. The electrical resistivity $\rho_{ ab}$ (parallel to the $ab$-plane) and specific heat $C$ were measured using a Quantum Design PPMS.

\begin{figure}[t]
\begin{center}
\includegraphics[width=7cm]{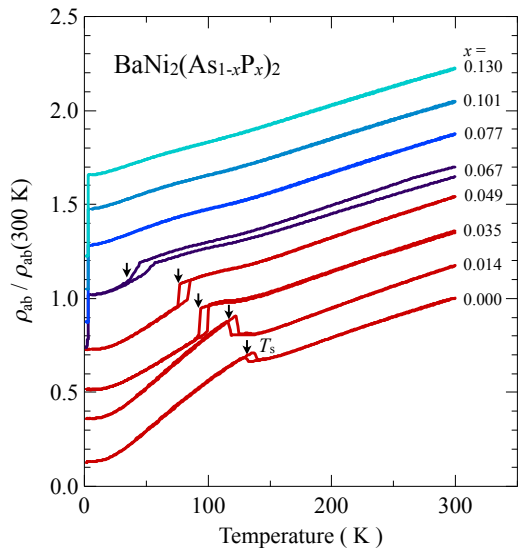}
\caption{
(Color online)
Temperature dependence of the electrical resistivity  parallel to the $ab$ plane, $\rho_{\rm ab}$, normalized by the value at 300 K for BaNi$_2$(As$_{1-x}$P$_x$)$_2$.
The data measured upon heating and cooling are plotted.
For the sake of clarity, $\rho_{\rm ab}/\rho_{\rm ab}$(300 K) is shifted by 0.175 with respect to all data.
$T_{\rm s}$ is the phase transition temperature at which the tetragonal-to-triclinic phase transition occurs; it is determined as the midpoint of the jump in $\rho_{\rm ab}$.
The residual resistivity $\rho_0$ ranges from 10 to 30 $\mu\Omega$cm.
}
\end{center}
\end{figure}

\begin{figure}[t]
\begin{center}
\includegraphics[width=7cm]{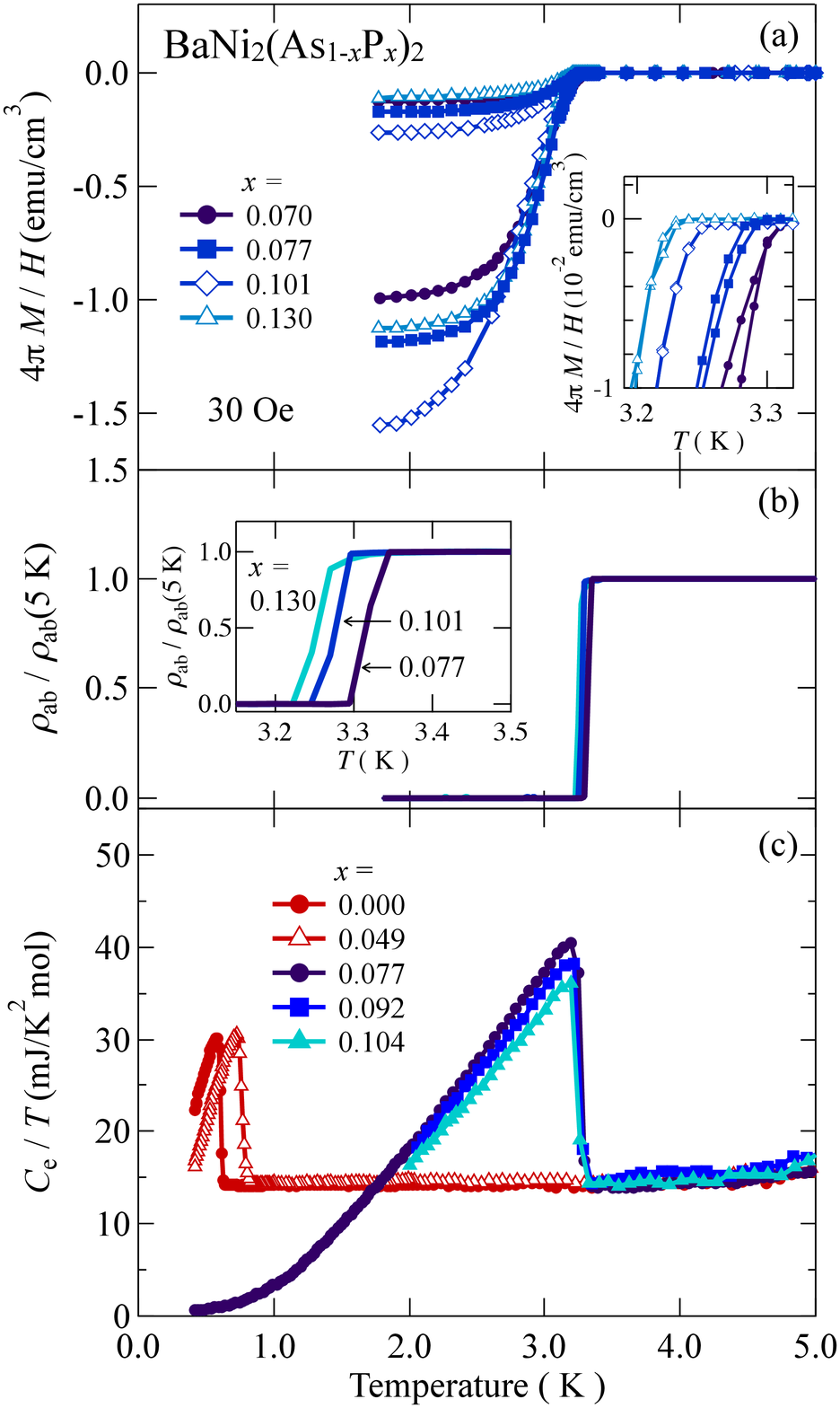}
\caption{
(Color online)
(a) Temperature dependences of dc magnetization $M$ measured in a magnetic field $H$ of 30 Oe for BaNi$_2$(As$_{1-x}$P$_x$)$_2$ under zero-field-cooling (ZFC) and field-cooling (FC).
(b) Temperature dependence of the electrical resistivity parallel to the $ab$ plane, $\rho_{\rm ab}$, normalized by the value at 5 K for BaNi$_2$(As$_{1-x}$P$_x$)$_2$.
(c) Temperature dependence of the electronic specific heat divided by the temperature, $C_e/T$, for BaNi$_2$(As$_{1-x}$P$_x$)$_2$.
$C_e$ was determined by subtracting the phonon contribution $\beta T^3$ from the total specific heat $C$, as shown in Fig.~3. The insets show magnified views of the vicinity of the superconducting transition.
}
\end{center}
\end{figure}

Figure 1 shows the temperature dependence of the electrical resistivity for BaNi$_2$(As$_{1-x}$P$_x$)$_2$. In a manner consistent with previous reports, pure BaNi$_2$As$_2$ exhibits a transition at 130 K with a thermal hysteresis accompanying a sudden increase in resistivity upon cooling \cite{0953-8984-20-34-342203,PhysRevB.79.094508,PhysRevLett.102.147004}.
For 3.5\% phosphorous doping, the transition is significantly suppressed to 90$-$100 K and the sudden resistivity increase changes into a sudden decrease upon cooling. The transition appears to be absent for 7.7\% phosphorous doping. These results suggest the suppression of the triclinic phase at $x$ $\simeq$ 0.07 in BaNi$_2$(As$_{1-x}$P$_x$)$_2$.

We found that superconductivity emerges below 3.3 K as soon as the triclinic phase is suppressed with phosphorous doping, while it emerges below 0.7 K in the triclinic phase at $x$ $<$ 0.07. This is demonstrated by the low-temperature magnetic susceptibility and resistivity data shown in Fig.~2. For $x$ = 0.070, the bulk superconductivity is evident from the full shielding diamagnetic signal and sharp resistivity transition at 3.33 K. The superconductivity persists until $x$ reaches the solubility limit at $x$ = 0.13, while $T_c$ decreases slightly to 3.24 K.
The low-temperature specific-heat data, shown in Figs.~2(c) and 3, give further evidence of the enhanced superconductivity in the tetragonal phase. Pure BaNi$_2$As$_2$ exhibits a specific heat jump at 0.6 K, as reported previously \cite{0953-8984-20-34-342203,PhysRevB.79.094508,PhysRevLett.102.147004}.
The transition temperature increases slightly with phosphorous doping, but remains below 0.97 K as long as the system is in the triclinic phase. 
In the tetragonal phase with more than 7.7\% phosphorous doping, the specific heat jump appears at an elevated temperature of 3.3 K, in a manner consistent with the magnetic and resistivity data.

\begin{figure}[t]
\begin{center}
\includegraphics[width=6cm]{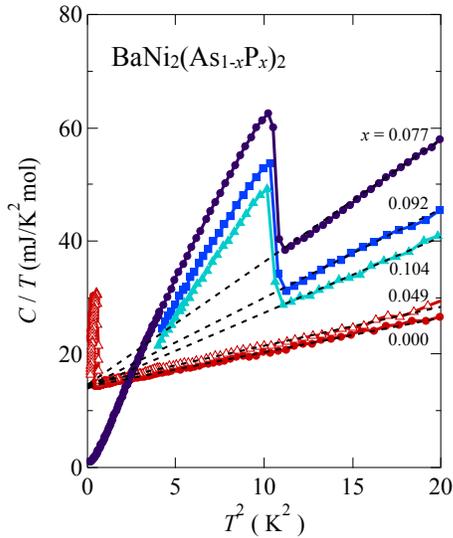}
\caption{
(Color online) The specific heat divided by the temperature, $C/T$, as a function of $T^2$ for BaNi$_2$(As$_{1-x}$P$_x$)$_2$. The broken lines denote the fits by $C/T = \gamma + \beta T^2$, where $\gamma$ is the electronic specific heat coefficient and $\beta$, a constant corresponding to the Debye phonon contributions.
}
\end{center}
\end{figure}

\begin{figure}[t]
\begin{center}
\includegraphics[width=8cm]{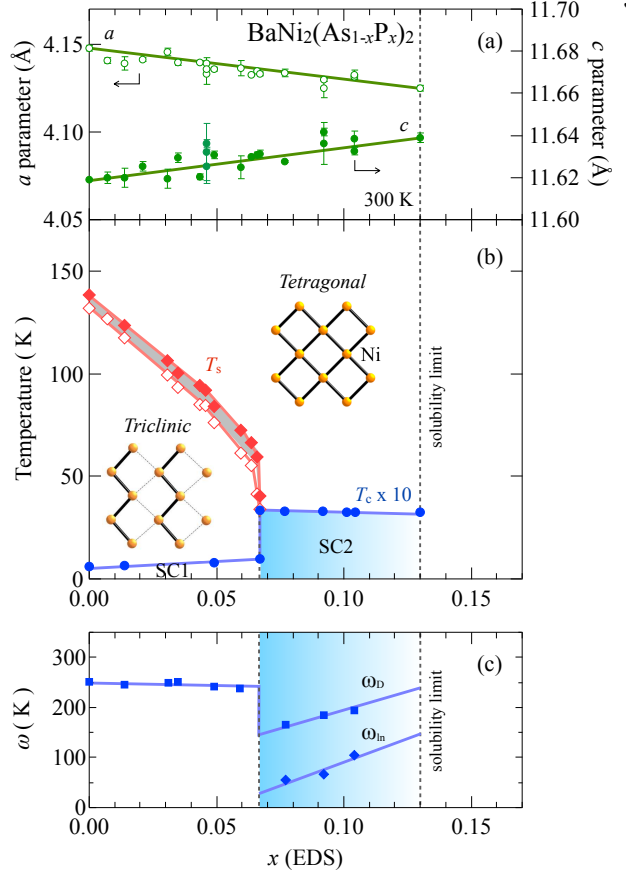}
\caption{
(Color online)
(a) Lattice parameters $a$ and $c$ as a function of phosphorous content $x$ at 300 K for BaNi$_2$(As$_{1-x}$P$_x$)$_2$.
(b) Electronic phase diagram of BaNi$_2$(As$_{1-x}$P$_x$)$_2$. The (blue) closed circles represent the superconducting transition temperatures $T_{c}$. 
For clarity, the values of $T_{\rm c}$ have been scaled by a factor of ten. 
SC1 and SC2 denote the superconducting phases. The (red) open and closed diamonds represent the tetragonal-to-triclinic structural transition temperatures $T_s$ upon cooling and heating, respectively. 
The insets show schematic views of Ni planes in the triclinic and tetragonal phase.
(c) The Debye frequency $\omega_{\rm D}$ and the logarithmic-averaged phonon frequency $\omega_{\rm ln}$ as a function of phosphorous content $x$ for BaNi$_2$(As$_{1-x}$P$_x$)$_2$.
$\omega_{\rm ln}$ was determined from the normalized specific-heat jump in Fig.~2(c) and $\omega_{\rm D}$, from the slope of the $C/T$ vs. $T^2$ curves in Fig.~3.
}
\end{center}
\end{figure}

The enhanced superconductivity at the structural phase boundary in phosphorus-doped BaNi$_2$As$_2$ is reminiscent of tellurium at the $\beta$-Po bcc phase boundary under high pressure \cite{PhysRevLett.44.1623,Akahama1992803,PhysRevB.65.064504,PhysRevLett.77.1151,PhysRevB.63.224107,0953-8984-19-12-125206}. 
Mauri {\it et al.} proposed that the enhanced superconductivity in Te is due to the softening of transverse phonons along the $\Gamma$N line in the bcc phase and the resultant enhancement of electron-phonon coupling under pressure \cite{PhysRevLett.77.1151}.
In order to examine the role of soft phonons in the enhanced $T_c$ of phosphorus-doped BaNi$_2$As$_2$, the normal-state specific heat was examined at low temperatures. Figure 3 shows the specific heat divided by the temperature $C/T$ as a function of the squared temperature $T^2$. The normal-state data above $T_c$ can be well fitted by $C/T = \gamma + \beta T^2$, where $\gamma$ is an electronic specific-heat coefficient and $\beta$, the coefficient of phonon contributions from which the Debye frequency $\omega_{\rm D}$ is estimated.
As can be seen from Fig.~3, the slope of the $C/T$ vs. $T^2$ lines is almost unchanged in the triclinic phase for less than 7\% phosphorus doping. The slope increases suddenly for 7.7\% phosphorous doping, suggesting the occurrence of significant phonon softening when the triclinic-to-tetragonal phase transition occurs. The slope decreases upon further phosphorous doping in the tetragonal phase.
Figure 4(c) shows the estimated Debye frequency $\omega_{\rm D}$ as a function of the phosphorus content $x$.
$\omega_{\rm D}$ shows a significant reduction from 250 to 150 K at the structural phase boundary of $x$ = 0.07.

In contrast to the strong dependence of both $\omega_{\rm D}$ and $T_c$ on doping, the electronic specific-heat coefficient $\gamma$ ($\simeq$ 14 mJ/molK$^{2}$) is almost independent of the phosphorus content $x$, as can be seen from the almost unchanged intercept of the $C/T$ vs. $T^2$ lines along the $C/T$ axis in Fig.~3. Furthermore, the electronic density of states at the Fermi level $\simeq$ 6 eV$^{-1}$ per formula unit, as determined from the specific heat $\gamma$, is comparable to the band calculation value 3.57 eV$^{-1}$ for the tetragonal phase of pure BaNi$_2$As$_2$ \cite{PhysRevB.78.132511}.
These observations suggest that the structural phase transition, as well as the enhanced superconductivity in the tetragonal phase, is not electronic but phonon, specifically, enhanced electron-phonon coupling due to soft phonons, in origin. 
Indeed, the normalized specific heat jump $\Delta C / \gamma T_{\rm c}$ $\simeq$ 1.3 for the triclinic phase at $x$ $<$ 0.07, determined from the data shown in Fig.~2(c), is comparable to the value of the weak coupling limit (=1.43), while $\Delta C / \gamma T_{\rm c}$ is enhanced in the tetragonal phase with a maximum value of 1.90 at $x$ = 0.077, indicative of strong coupling superconductivity. $\Delta C / \gamma T_{\rm c}$ is gradually decreased upon further doping to 1.78 and 1.60 for $x =$ 0.092 and 0.104, respectively, as can be seen from Fig.~2(c) \cite{gamma}. 
We estimate the logarithmic-averaged phonon frequency $\omega_{\rm ln}$ using the relationship for strong-coupling superconductors $\Delta C(T_{\rm c})/\gamma T_{\rm c} = 1.43[1 + (53/x^2){\rm ln}(x/3)]$, where $x = \omega_{\rm ln}/T_{\rm c}$ \cite{RMP.62.1027,PhysRevB.76.014523,JPSJ.78.064702}.
Figure 4(c) shows $\omega_{\rm ln}$, together with the Debye frequency $\omega_{\rm D}$, as a function of the phosphorus content $x$. Both $\omega_{\rm ln}$ and $\omega_{\rm D}$ exhibit significant softening as the system approaches the structural phase boundary from the tetragonal side, suggesting that the low-lying soft phonons, which are coupled strongly with acoustic modes, play an important role in the emergence of the strong-coupling superconductivity in the tetragonal phase.
The in-plane Ni and As mode at $\sim$ 50 cm$^{-1}$ (equivalent to 80 K) \cite{PhysRevB.78.132511}, which is thought to be quenched to form alternating Ni-As bonds in the triclinic phase, could be a soft mode, although further experimental confirmations of the same are required.

Our results are summarized in the electronic phase diagram shown in Fig.~4(b). The triclinic phase transition temperature in pure BaNi$_2$As$_2$ is strongly suppressed with phosphorus doping. As a result, the superconducting transition temperature is enhanced from $T_c$ = 0.6$-$0.9 K in the triclinic phase to 3.2$-$3.3 K in the tetragonal phase.
A similar step-wise shape of the superconducting phase boundary, where another structural phase boundary intersects, has been observed in tellurium under high pressure \cite{PhysRevLett.77.1151}. 
A similar mechanism may be active in the phosphorous-doped BaNi$_2$As$_2$. Phosphorous is isovalent with arsenic, but has a smaller ionic radius. Therefore, phosphorous doping serves to create chemical pressure. Indeed, lattice parameters $a$ and $c$ at 300 K show monotonic dependence on the phosphorus content $x$, as shown in Fig.~4(a); thus, the unit cell volume decreases monotonically from 199.9 \AA$^3$ for pure BaNi$_2$As$_2$ to 198.1 \AA$^3$ for 13\% phosphorus doping of BaNi$_2$(As$_{1-x}$P$_x$)$_2$.
Soft phonons in the tetragonal phase, indicated by both $\omega_{\rm D}$ and $\omega_{\rm ln}$, shown in Fig.~4(c), could enhance the electron-phonon coupling and thus the superconductivity.
When the phosphorus content $x$ is decreased from the tetragonal side, the soft phonon leads to a structural instability and eventually results in the triclinic phase transition and suppressed superconductivity.
There exists a miscibility gap of the phosphorus content above $x$ $=$ 0.13. It should be noted that superconductivity below $T_c$ = 3 K appears when arsenic is completely replaced by phosphorus in BaNi$_2$P$_2$  \cite{SSC.147.111}.

To conclude, our experiments show that the enhancement of superconductivity associated with phonon softening and subsequent structural phase transition occurs because of the phosphorus doping of BaNi$_2$As$_2$. Specifically, the increase in $T_c$ from 0.6 to 3.3 K (approximately five-fold enhancement) is related to the giant phonon softening of approximately 50\% at the triclinic-to-tetragonal phase transition induced by phosphorus doping. Our results demonstrate that the chemical tuning of soft phonons is practical and effective for optimizing superconductivity.

Part of this work was performed at the Advanced Science Research Center, Okayama University. It was partially supported by a Grant-in-Aid for Young Scientists (B) (23740274) from the Japan Society for the Promotion of Science (JSPS) and the Funding Program for World-Leading Innovative R\&D on Science and Technology (FIRST Program) from the JSPS.


\begin{thebibliography}{26}%
\makeatletter
\providecommand \@ifxundefined [1]{%
 \@ifx{#1\undefined}
}%
\providecommand \@ifnum [1]{%
 \ifnum #1\expandafter \@firstoftwo
 \else \expandafter \@secondoftwo
 \fi
}%
\providecommand \@ifx [1]{%
 \ifx #1\expandafter \@firstoftwo
 \else \expandafter \@secondoftwo
 \fi
}%
\providecommand \natexlab [1]{#1}%
\providecommand \enquote  [1]{``#1''}%
\providecommand \bibnamefont  [1]{#1}%
\providecommand \bibfnamefont [1]{#1}%
\providecommand \citenamefont [1]{#1}%
\providecommand \href@noop [0]{\@secondoftwo}%
\providecommand \href [0]{\begingroup \@sanitize@url \@href}%
\providecommand \@href[1]{\@@startlink{#1}\@@href}%
\providecommand \@@href[1]{\endgroup#1\@@endlink}%
\providecommand \@sanitize@url [0]{\catcode `\\12\catcode `\$12\catcode
  `\&12\catcode `\#12\catcode `\^12\catcode `\_12\catcode `\%12\relax}%
\providecommand \@@startlink[1]{}%
\providecommand \@@endlink[0]{}%
\providecommand \url  [0]{\begingroup\@sanitize@url \@url }%
\providecommand \@url [1]{\endgroup\@href {#1}{\urlprefix }}%
\providecommand \urlprefix  [0]{URL }%
\providecommand \Eprint [0]{\href }%
\providecommand \doibase [0]{http://dx.doi.org/}%
\providecommand \selectlanguage [0]{\@gobble}%
\providecommand \bibinfo  [0]{\@secondoftwo}%
\providecommand \bibfield  [0]{\@secondoftwo}%
\providecommand \translation [1]{[#1]}%
\providecommand \BibitemOpen [0]{}%
\providecommand \bibitemStop [0]{}%
\providecommand \bibitemNoStop [0]{.\EOS\space}%
\providecommand \EOS [0]{\spacefactor3000\relax}%
\providecommand \BibitemShut  [1]{\csname bibitem#1\endcsname}%
\let\auto@bib@innerbib\@empty
\bibitem [{\citenamefont {d'Astuto}\ \emph {et~al.}(2002)\citenamefont
  {d'Astuto}, \citenamefont {Mang}, \citenamefont {Giura}, \citenamefont
  {Shukla}, \citenamefont {Ghigna}, \citenamefont {Mirone}, \citenamefont
  {Braden}, \citenamefont {Greven}, \citenamefont {Krisch},\ and\ \citenamefont
  {Sette}}]{PhysRevLett.88.167002}%
  \BibitemOpen
  \bibfield  {author} {\bibinfo {author} {\bibfnamefont {M.}~\bibnamefont
  {d'Astuto}}, \bibinfo {author} {\bibfnamefont {P.~K.}\ \bibnamefont {Mang}},
  \bibinfo {author} {\bibfnamefont {P.}~\bibnamefont {Giura}}, \bibinfo
  {author} {\bibfnamefont {A.}~\bibnamefont {Shukla}}, \bibinfo {author}
  {\bibfnamefont {P.}~\bibnamefont {Ghigna}}, \bibinfo {author} {\bibfnamefont
  {A.}~\bibnamefont {Mirone}}, \bibinfo {author} {\bibfnamefont
  {M.}~\bibnamefont {Braden}}, \bibinfo {author} {\bibfnamefont
  {M.}~\bibnamefont {Greven}}, \bibinfo {author} {\bibfnamefont
  {M.}~\bibnamefont {Krisch}}, \ and\ \bibinfo {author} {\bibfnamefont
  {F.}~\bibnamefont {Sette}},\ }\href {\doibase 10.1103/PhysRevLett.88.167002}
  {\bibfield  {journal} {\bibinfo  {journal} {Phys. Rev. Lett.}\ }\textbf
  {\bibinfo {volume} {88}},\ \bibinfo {pages} {167002} (\bibinfo {year}
  {2002})}\BibitemShut {NoStop}%
\bibitem [{\citenamefont {Reznik}\ \emph {et~al.}(2006)\citenamefont {Reznik},
  \citenamefont {Pintschovius}, \citenamefont {Ito}, \citenamefont {Iikubo},
  \citenamefont {Sato}, \citenamefont {Goka}, \citenamefont {Fujita},
  \citenamefont {Yamada}, \citenamefont {Gu},\ and\ \citenamefont
  {Tranquada}}]{Nature04704}%
  \BibitemOpen
  \bibfield  {author} {\bibinfo {author} {\bibfnamefont {D.}~\bibnamefont
  {Reznik}}, \bibinfo {author} {\bibfnamefont {L.}~\bibnamefont
  {Pintschovius}}, \bibinfo {author} {\bibfnamefont {M.}~\bibnamefont {Ito}},
  \bibinfo {author} {\bibfnamefont {S.}~\bibnamefont {Iikubo}}, \bibinfo
  {author} {\bibfnamefont {M.}~\bibnamefont {Sato}}, \bibinfo {author}
  {\bibfnamefont {H.}~\bibnamefont {Goka}}, \bibinfo {author} {\bibfnamefont
  {M.}~\bibnamefont {Fujita}}, \bibinfo {author} {\bibfnamefont
  {K.}~\bibnamefont {Yamada}}, \bibinfo {author} {\bibfnamefont {G.~D.}\
  \bibnamefont {Gu}}, \ and\ \bibinfo {author} {\bibfnamefont {J.~M.}\
  \bibnamefont {Tranquada}},\ }\href {\doibase 10.1038/nature04704} {\bibfield
  {journal} {\bibinfo  {journal} {Nature}\ }\textbf {\bibinfo {volume} {440}},\
  \bibinfo {pages} {1170} (\bibinfo {year} {2006})}\BibitemShut {NoStop}%
\bibitem [{\citenamefont {Goto}\ \emph {et~al.}(2011)\citenamefont {Goto},
  \citenamefont {Kurihara}, \citenamefont {Araki}, \citenamefont {Mitsumoto},
  \citenamefont {Akatsu}, \citenamefont {Nemoto}, \citenamefont {Tatematsu},\
  and\ \citenamefont {Sato}}]{JPSJ.80.073702}%
  \BibitemOpen
  \bibfield  {author} {\bibinfo {author} {\bibfnamefont {T.}~\bibnamefont
  {Goto}}, \bibinfo {author} {\bibfnamefont {R.}~\bibnamefont {Kurihara}},
  \bibinfo {author} {\bibfnamefont {K.}~\bibnamefont {Araki}}, \bibinfo
  {author} {\bibfnamefont {K.}~\bibnamefont {Mitsumoto}}, \bibinfo {author}
  {\bibfnamefont {M.}~\bibnamefont {Akatsu}}, \bibinfo {author} {\bibfnamefont
  {Y.}~\bibnamefont {Nemoto}}, \bibinfo {author} {\bibfnamefont
  {S.}~\bibnamefont {Tatematsu}}, \ and\ \bibinfo {author} {\bibfnamefont
  {M.}~\bibnamefont {Sato}},\ }\href {\doibase 10.1143/JPSJ.80.073702}
  {\bibfield  {journal} {\bibinfo  {journal} {J. Phys. Soc. Jpn.}\ }\textbf
  {\bibinfo {volume} {80}},\ \bibinfo {pages} {073702} (\bibinfo {year}
  {2011})}\BibitemShut {NoStop}%
\bibitem [{\citenamefont {Yoshizawa}\ \emph {et~al.}(2012)\citenamefont
  {Yoshizawa}, \citenamefont {Kimura}, \citenamefont {Chiba}, \citenamefont
  {Simayi}, \citenamefont {Nakanishi}, \citenamefont {Kihou}, \citenamefont
  {Lee}, \citenamefont {Iyo}, \citenamefont {Eisaki}, \citenamefont
  {Nakajima},\ and\ \citenamefont {Uchida}}]{JPSJ.81.024604}%
  \BibitemOpen
  \bibfield  {author} {\bibinfo {author} {\bibfnamefont {M.}~\bibnamefont
  {Yoshizawa}}, \bibinfo {author} {\bibfnamefont {D.}~\bibnamefont {Kimura}},
  \bibinfo {author} {\bibfnamefont {T.}~\bibnamefont {Chiba}}, \bibinfo
  {author} {\bibfnamefont {S.}~\bibnamefont {Simayi}}, \bibinfo {author}
  {\bibfnamefont {Y.}~\bibnamefont {Nakanishi}}, \bibinfo {author}
  {\bibfnamefont {K.}~\bibnamefont {Kihou}}, \bibinfo {author} {\bibfnamefont
  {C.-H.}\ \bibnamefont {Lee}}, \bibinfo {author} {\bibfnamefont
  {A.}~\bibnamefont {Iyo}}, \bibinfo {author} {\bibfnamefont {H.}~\bibnamefont
  {Eisaki}}, \bibinfo {author} {\bibfnamefont {M.}~\bibnamefont {Nakajima}}, \
  and\ \bibinfo {author} {\bibfnamefont {S.}~\bibnamefont {Uchida}},\ }\href
  {\doibase 10.1143/JPSJ.81.024604} {\bibfield  {journal} {\bibinfo  {journal}
  {J. Phys. Soc. Jpn.}\ }\textbf {\bibinfo {volume} {81}},\ \bibinfo {pages}
  {024604} (\bibinfo {year} {2012})}\BibitemShut {NoStop}%
\bibitem [{\citenamefont {Testardi}(1975)}]{RevModPhys.47.637}%
  \BibitemOpen
  \bibfield  {author} {\bibinfo {author} {\bibfnamefont {L.~R.}\ \bibnamefont
  {Testardi}},\ }\href {\doibase 10.1103/RevModPhys.47.637} {\bibfield
  {journal} {\bibinfo  {journal} {Rev. Mod. Phys.}\ }\textbf {\bibinfo {volume}
  {47}},\ \bibinfo {pages} {637} (\bibinfo {year} {1975})}\BibitemShut
  {NoStop}%
\bibitem [{\citenamefont {Smith}\ \emph {et~al.}(2006)\citenamefont {Smith},
  \citenamefont {Kusmartseva}, \citenamefont {Ko}, \citenamefont {Saxena},
  \citenamefont {Akrap}, \citenamefont {Forr\'o}, \citenamefont {Laad},
  \citenamefont {Weller}, \citenamefont {Ellerby},\ and\ \citenamefont
  {Skipper}}]{PhysRevB.74.024505}%
  \BibitemOpen
  \bibfield  {author} {\bibinfo {author} {\bibfnamefont {R.~P.}\ \bibnamefont
  {Smith}}, \bibinfo {author} {\bibfnamefont {A.}~\bibnamefont {Kusmartseva}},
  \bibinfo {author} {\bibfnamefont {Y.~T.~C.}\ \bibnamefont {Ko}}, \bibinfo
  {author} {\bibfnamefont {S.~S.}\ \bibnamefont {Saxena}}, \bibinfo {author}
  {\bibfnamefont {A.}~\bibnamefont {Akrap}}, \bibinfo {author} {\bibfnamefont
  {L.}~\bibnamefont {Forr\'o}}, \bibinfo {author} {\bibfnamefont
  {M.}~\bibnamefont {Laad}}, \bibinfo {author} {\bibfnamefont {T.~E.}\
  \bibnamefont {Weller}}, \bibinfo {author} {\bibfnamefont {M.}~\bibnamefont
  {Ellerby}}, \ and\ \bibinfo {author} {\bibfnamefont {N.~T.}\ \bibnamefont
  {Skipper}},\ }\href {\doibase 10.1103/PhysRevB.74.024505} {\bibfield
  {journal} {\bibinfo  {journal} {Phys. Rev. B}\ }\textbf {\bibinfo {volume}
  {74}},\ \bibinfo {pages} {024505} (\bibinfo {year} {2006})}\BibitemShut
  {NoStop}%
\bibitem [{\citenamefont {Kim}\ \emph {et~al.}(2006)\citenamefont {Kim},
  \citenamefont {Boeri}, \citenamefont {Kremer},\ and\ \citenamefont
  {Razavi}}]{PhysRevB.74.214513}%
  \BibitemOpen
  \bibfield  {author} {\bibinfo {author} {\bibfnamefont {J.~S.}\ \bibnamefont
  {Kim}}, \bibinfo {author} {\bibfnamefont {L.}~\bibnamefont {Boeri}}, \bibinfo
  {author} {\bibfnamefont {R.~K.}\ \bibnamefont {Kremer}}, \ and\ \bibinfo
  {author} {\bibfnamefont {F.~S.}\ \bibnamefont {Razavi}},\ }\href {\doibase
  10.1103/PhysRevB.74.214513} {\bibfield  {journal} {\bibinfo  {journal} {Phys.
  Rev. B}\ }\textbf {\bibinfo {volume} {74}},\ \bibinfo {pages} {214513}
  (\bibinfo {year} {2006})}\BibitemShut {NoStop}%
\bibitem [{\citenamefont {Gauzzi}\ \emph {et~al.}(2007)\citenamefont {Gauzzi},
  \citenamefont {Takashima}, \citenamefont {Takeshita}, \citenamefont
  {Terakura}, \citenamefont {Takagi}, \citenamefont {Emery}, \citenamefont
  {H\'erold}, \citenamefont {Lagrange},\ and\ \citenamefont
  {Loupias}}]{PhysRevLett.98.067002}%
  \BibitemOpen
  \bibfield  {author} {\bibinfo {author} {\bibfnamefont {A.}~\bibnamefont
  {Gauzzi}}, \bibinfo {author} {\bibfnamefont {S.}~\bibnamefont {Takashima}},
  \bibinfo {author} {\bibfnamefont {N.}~\bibnamefont {Takeshita}}, \bibinfo
  {author} {\bibfnamefont {C.}~\bibnamefont {Terakura}}, \bibinfo {author}
  {\bibfnamefont {H.}~\bibnamefont {Takagi}}, \bibinfo {author} {\bibfnamefont
  {N.}~\bibnamefont {Emery}}, \bibinfo {author} {\bibfnamefont
  {C.}~\bibnamefont {H\'erold}}, \bibinfo {author} {\bibfnamefont
  {P.}~\bibnamefont {Lagrange}}, \ and\ \bibinfo {author} {\bibfnamefont
  {G.}~\bibnamefont {Loupias}},\ }\href {\doibase
  10.1103/PhysRevLett.98.067002} {\bibfield  {journal} {\bibinfo  {journal}
  {Phys. Rev. Lett.}\ }\textbf {\bibinfo {volume} {98}},\ \bibinfo {pages}
  {067002} (\bibinfo {year} {2007})}\BibitemShut {NoStop}%
\bibitem [{\citenamefont {Gauzzi}\ \emph {et~al.}(2008)\citenamefont {Gauzzi},
  \citenamefont {Bendiab}, \citenamefont {d'Astuto}, \citenamefont {Canny},
  \citenamefont {Calandra}, \citenamefont {Mauri}, \citenamefont {Loupias},
  \citenamefont {Emery}, \citenamefont {H\'erold}, \citenamefont {Lagrange},
  \citenamefont {Hanfland},\ and\ \citenamefont
  {Mezouar}}]{PhysRevB.78.064506}%
  \BibitemOpen
  \bibfield  {author} {\bibinfo {author} {\bibfnamefont {A.}~\bibnamefont
  {Gauzzi}}, \bibinfo {author} {\bibfnamefont {N.}~\bibnamefont {Bendiab}},
  \bibinfo {author} {\bibfnamefont {M.}~\bibnamefont {d'Astuto}}, \bibinfo
  {author} {\bibfnamefont {B.}~\bibnamefont {Canny}}, \bibinfo {author}
  {\bibfnamefont {M.}~\bibnamefont {Calandra}}, \bibinfo {author}
  {\bibfnamefont {F.}~\bibnamefont {Mauri}}, \bibinfo {author} {\bibfnamefont
  {G.}~\bibnamefont {Loupias}}, \bibinfo {author} {\bibfnamefont
  {N.}~\bibnamefont {Emery}}, \bibinfo {author} {\bibfnamefont
  {C.}~\bibnamefont {H\'erold}}, \bibinfo {author} {\bibfnamefont
  {P.}~\bibnamefont {Lagrange}}, \bibinfo {author} {\bibfnamefont
  {M.}~\bibnamefont {Hanfland}}, \ and\ \bibinfo {author} {\bibfnamefont
  {M.}~\bibnamefont {Mezouar}},\ }\href {\doibase 10.1103/PhysRevB.78.064506}
  {\bibfield  {journal} {\bibinfo  {journal} {Phys. Rev. B}\ }\textbf {\bibinfo
  {volume} {78}},\ \bibinfo {pages} {064506} (\bibinfo {year}
  {2008})}\BibitemShut {NoStop}%
\bibitem [{\citenamefont {Profeta}\ \emph {et~al.}(2006)\citenamefont
  {Profeta}, \citenamefont {Franchini}, \citenamefont {Lathiotakis},
  \citenamefont {Floris}, \citenamefont {Sanna}, \citenamefont {Marques},
  \citenamefont {L\"uders}, \citenamefont {Massidda}, \citenamefont {Gross},\
  and\ \citenamefont {Continenza}}]{PhysRevLett.96.047003}%
  \BibitemOpen
  \bibfield  {author} {\bibinfo {author} {\bibfnamefont {G.}~\bibnamefont
  {Profeta}}, \bibinfo {author} {\bibfnamefont {C.}~\bibnamefont {Franchini}},
  \bibinfo {author} {\bibfnamefont {N.~N.}\ \bibnamefont {Lathiotakis}},
  \bibinfo {author} {\bibfnamefont {A.}~\bibnamefont {Floris}}, \bibinfo
  {author} {\bibfnamefont {A.}~\bibnamefont {Sanna}}, \bibinfo {author}
  {\bibfnamefont {M.~A.~L.}\ \bibnamefont {Marques}}, \bibinfo {author}
  {\bibfnamefont {M.}~\bibnamefont {L\"uders}}, \bibinfo {author}
  {\bibfnamefont {S.}~\bibnamefont {Massidda}}, \bibinfo {author}
  {\bibfnamefont {E.~K.~U.}\ \bibnamefont {Gross}}, \ and\ \bibinfo {author}
  {\bibfnamefont {A.}~\bibnamefont {Continenza}},\ }\href {\doibase
  10.1103/PhysRevLett.96.047003} {\bibfield  {journal} {\bibinfo  {journal}
  {Phys. Rev. Lett.}\ }\textbf {\bibinfo {volume} {96}},\ \bibinfo {pages}
  {047003} (\bibinfo {year} {2006})}\BibitemShut {NoStop}%
\bibitem [{\citenamefont {Bundy}\ and\ \citenamefont
  {Dunn}(1980)}]{PhysRevLett.44.1623}%
  \BibitemOpen
  \bibfield  {author} {\bibinfo {author} {\bibfnamefont {F.~P.}\ \bibnamefont
  {Bundy}}\ and\ \bibinfo {author} {\bibfnamefont {K.~J.}\ \bibnamefont
  {Dunn}},\ }\href {\doibase 10.1103/PhysRevLett.44.1623} {\bibfield  {journal}
  {\bibinfo  {journal} {Phys. Rev. Lett.}\ }\textbf {\bibinfo {volume} {44}},\
  \bibinfo {pages} {1623} (\bibinfo {year} {1980})}\BibitemShut {NoStop}%
\bibitem [{\citenamefont {Akahama}\ \emph {et~al.}(1992)\citenamefont
  {Akahama}, \citenamefont {Kobayashi},\ and\ \citenamefont
  {Kawamura}}]{Akahama1992803}%
  \BibitemOpen
  \bibfield  {author} {\bibinfo {author} {\bibfnamefont {Y.}~\bibnamefont
  {Akahama}}, \bibinfo {author} {\bibfnamefont {M.}~\bibnamefont {Kobayashi}},
  \ and\ \bibinfo {author} {\bibfnamefont {H.}~\bibnamefont {Kawamura}},\
  }\href {\doibase 10.1016/0038-1098(92)90093-O} {\bibfield  {journal}
  {\bibinfo  {journal} {Solid State Commun.}\ }\textbf {\bibinfo {volume}
  {84}},\ \bibinfo {pages} {803 } (\bibinfo {year} {1992})}\BibitemShut
  {NoStop}%
\bibitem [{\citenamefont {Gregoryanz}\ \emph {et~al.}(2002)\citenamefont
  {Gregoryanz}, \citenamefont {Struzhkin}, \citenamefont {Hemley},
  \citenamefont {Eremets}, \citenamefont {Mao},\ and\ \citenamefont
  {Timofeev}}]{PhysRevB.65.064504}%
  \BibitemOpen
  \bibfield  {author} {\bibinfo {author} {\bibfnamefont {E.}~\bibnamefont
  {Gregoryanz}}, \bibinfo {author} {\bibfnamefont {V.~V.}\ \bibnamefont
  {Struzhkin}}, \bibinfo {author} {\bibfnamefont {R.~J.}\ \bibnamefont
  {Hemley}}, \bibinfo {author} {\bibfnamefont {M.~I.}\ \bibnamefont {Eremets}},
  \bibinfo {author} {\bibfnamefont {H.-k.}\ \bibnamefont {Mao}}, \ and\
  \bibinfo {author} {\bibfnamefont {Y.~A.}\ \bibnamefont {Timofeev}},\ }\href
  {\doibase 10.1103/PhysRevB.65.064504} {\bibfield  {journal} {\bibinfo
  {journal} {Phys. Rev. B}\ }\textbf {\bibinfo {volume} {65}},\ \bibinfo
  {pages} {064504} (\bibinfo {year} {2002})}\BibitemShut {NoStop}%
\bibitem [{\citenamefont {Mauri}\ \emph {et~al.}(1996)\citenamefont {Mauri},
  \citenamefont {Zakharov}, \citenamefont {de~Gironcoli}, \citenamefont
  {Louie},\ and\ \citenamefont {Cohen}}]{PhysRevLett.77.1151}%
  \BibitemOpen
  \bibfield  {author} {\bibinfo {author} {\bibfnamefont {F.}~\bibnamefont
  {Mauri}}, \bibinfo {author} {\bibfnamefont {O.}~\bibnamefont {Zakharov}},
  \bibinfo {author} {\bibfnamefont {S.}~\bibnamefont {de~Gironcoli}}, \bibinfo
  {author} {\bibfnamefont {S.~G.}\ \bibnamefont {Louie}}, \ and\ \bibinfo
  {author} {\bibfnamefont {M.~L.}\ \bibnamefont {Cohen}},\ }\href {\doibase
  10.1103/PhysRevLett.77.1151} {\bibfield  {journal} {\bibinfo  {journal}
  {Phys. Rev. Lett.}\ }\textbf {\bibinfo {volume} {77}},\ \bibinfo {pages}
  {1151} (\bibinfo {year} {1996})}\BibitemShut {NoStop}%
\bibitem [{\citenamefont {Rudin}\ \emph {et~al.}(2001)\citenamefont {Rudin},
  \citenamefont {Liu}, \citenamefont {Freericks},\ and\ \citenamefont
  {Quandt}}]{PhysRevB.63.224107}%
  \BibitemOpen
  \bibfield  {author} {\bibinfo {author} {\bibfnamefont {S.~P.}\ \bibnamefont
  {Rudin}}, \bibinfo {author} {\bibfnamefont {A.~Y.}\ \bibnamefont {Liu}},
  \bibinfo {author} {\bibfnamefont {J.~K.}\ \bibnamefont {Freericks}}, \ and\
  \bibinfo {author} {\bibfnamefont {A.}~\bibnamefont {Quandt}},\ }\href
  {\doibase 10.1103/PhysRevB.63.224107} {\bibfield  {journal} {\bibinfo
  {journal} {Phys. Rev. B}\ }\textbf {\bibinfo {volume} {63}},\ \bibinfo
  {pages} {224107} (\bibinfo {year} {2001})}\BibitemShut {NoStop}%
\bibitem [{\citenamefont {Suzuki}\ and\ \citenamefont
  {Otani}(2007)}]{0953-8984-19-12-125206}%
  \BibitemOpen
  \bibfield  {author} {\bibinfo {author} {\bibfnamefont {N.}~\bibnamefont
  {Suzuki}}\ and\ \bibinfo {author} {\bibfnamefont {M.}~\bibnamefont {Otani}},\
  }\href {http://stacks.iop.org/0953-8984/19/i=12/a=125206} {\bibfield
  {journal} {\bibinfo  {journal} {J. Phys.: Condens. Matter}\ }\textbf
  {\bibinfo {volume} {19}},\ \bibinfo {pages} {125206} (\bibinfo {year}
  {2007})}\BibitemShut {NoStop}%
\bibitem [{\citenamefont {Ronning}\ \emph {et~al.}(2008)\citenamefont
  {Ronning}, \citenamefont {Kurita}, \citenamefont {Bauer}, \citenamefont
  {Scott}, \citenamefont {Park}, \citenamefont {Klimczuk}, \citenamefont
  {Movshovich},\ and\ \citenamefont {Thompson}}]{0953-8984-20-34-342203}%
  \BibitemOpen
  \bibfield  {author} {\bibinfo {author} {\bibfnamefont {F.}~\bibnamefont
  {Ronning}}, \bibinfo {author} {\bibfnamefont {N.}~\bibnamefont {Kurita}},
  \bibinfo {author} {\bibfnamefont {E.~D.}\ \bibnamefont {Bauer}}, \bibinfo
  {author} {\bibfnamefont {B.~L.}\ \bibnamefont {Scott}}, \bibinfo {author}
  {\bibfnamefont {T.}~\bibnamefont {Park}}, \bibinfo {author} {\bibfnamefont
  {T.}~\bibnamefont {Klimczuk}}, \bibinfo {author} {\bibfnamefont
  {R.}~\bibnamefont {Movshovich}}, \ and\ \bibinfo {author} {\bibfnamefont
  {J.~D.}\ \bibnamefont {Thompson}},\ }\href
  {http://stacks.iop.org/0953-8984/20/i=34/a=342203} {\bibfield  {journal}
  {\bibinfo  {journal} {J. Phys.: Condens. Matter}\ }\textbf {\bibinfo {volume}
  {20}},\ \bibinfo {pages} {342203} (\bibinfo {year} {2008})}\BibitemShut
  {NoStop}%
\bibitem [{\citenamefont {Sefat}\ \emph {et~al.}(2009)\citenamefont {Sefat},
  \citenamefont {McGuire}, \citenamefont {Jin}, \citenamefont {Sales},
  \citenamefont {Mandrus}, \citenamefont {Ronning}, \citenamefont {Bauer},\
  and\ \citenamefont {Mozharivskyj}}]{PhysRevB.79.094508}%
  \BibitemOpen
  \bibfield  {author} {\bibinfo {author} {\bibfnamefont {A.~S.}\ \bibnamefont
  {Sefat}}, \bibinfo {author} {\bibfnamefont {M.~A.}\ \bibnamefont {McGuire}},
  \bibinfo {author} {\bibfnamefont {R.}~\bibnamefont {Jin}}, \bibinfo {author}
  {\bibfnamefont {B.~C.}\ \bibnamefont {Sales}}, \bibinfo {author}
  {\bibfnamefont {D.}~\bibnamefont {Mandrus}}, \bibinfo {author} {\bibfnamefont
  {F.}~\bibnamefont {Ronning}}, \bibinfo {author} {\bibfnamefont {E.~D.}\
  \bibnamefont {Bauer}}, \ and\ \bibinfo {author} {\bibfnamefont
  {Y.}~\bibnamefont {Mozharivskyj}},\ }\href {\doibase
  10.1103/PhysRevB.79.094508} {\bibfield  {journal} {\bibinfo  {journal} {Phys.
  Rev. B}\ }\textbf {\bibinfo {volume} {79}},\ \bibinfo {pages} {094508}
  (\bibinfo {year} {2009})}\BibitemShut {NoStop}%
\bibitem [{\citenamefont {Kurita}\ \emph {et~al.}(2009)\citenamefont {Kurita},
  \citenamefont {Ronning}, \citenamefont {Tokiwa}, \citenamefont {Bauer},
  \citenamefont {Subedi}, \citenamefont {Singh}, \citenamefont {Thompson},\
  and\ \citenamefont {Movshovich}}]{PhysRevLett.102.147004}%
  \BibitemOpen
  \bibfield  {author} {\bibinfo {author} {\bibfnamefont {N.}~\bibnamefont
  {Kurita}}, \bibinfo {author} {\bibfnamefont {F.}~\bibnamefont {Ronning}},
  \bibinfo {author} {\bibfnamefont {Y.}~\bibnamefont {Tokiwa}}, \bibinfo
  {author} {\bibfnamefont {E.~D.}\ \bibnamefont {Bauer}}, \bibinfo {author}
  {\bibfnamefont {A.}~\bibnamefont {Subedi}}, \bibinfo {author} {\bibfnamefont
  {D.~J.}\ \bibnamefont {Singh}}, \bibinfo {author} {\bibfnamefont {J.~D.}\
  \bibnamefont {Thompson}}, \ and\ \bibinfo {author} {\bibfnamefont
  {R.}~\bibnamefont {Movshovich}},\ }\href {\doibase
  10.1103/PhysRevLett.102.147004} {\bibfield  {journal} {\bibinfo  {journal}
  {Phys. Rev. Lett.}\ }\textbf {\bibinfo {volume} {102}},\ \bibinfo {pages}
  {147004} (\bibinfo {year} {2009})}\BibitemShut {NoStop}%
\bibitem [{\citenamefont {Subedi}\ and\ \citenamefont
  {Singh}(2008)}]{PhysRevB.78.132511}%
  \BibitemOpen
  \bibfield  {author} {\bibinfo {author} {\bibfnamefont {A.}~\bibnamefont
  {Subedi}}\ and\ \bibinfo {author} {\bibfnamefont {D.~J.}\ \bibnamefont
  {Singh}},\ }\href {\doibase 10.1103/PhysRevB.78.132511} {\bibfield  {journal}
  {\bibinfo  {journal} {Phys. Rev. B}\ }\textbf {\bibinfo {volume} {78}},\
  \bibinfo {pages} {132511} (\bibinfo {year} {2008})}\BibitemShut {NoStop}%
\bibitem [{\citenamefont {Shein}\ and\ \citenamefont
  {Ivanovskii}(2009)}]{PhysRevB.79.054510}%
  \BibitemOpen
  \bibfield  {author} {\bibinfo {author} {\bibfnamefont {I.~R.}\ \bibnamefont
  {Shein}}\ and\ \bibinfo {author} {\bibfnamefont {A.~L.}\ \bibnamefont
  {Ivanovskii}},\ }\href {\doibase 10.1103/PhysRevB.79.054510} {\bibfield
  {journal} {\bibinfo  {journal} {Phys. Rev. B}\ }\textbf {\bibinfo {volume}
  {79}},\ \bibinfo {pages} {054510} (\bibinfo {year} {2009})}\BibitemShut
  {NoStop}%
\bibitem [{gam()}]{gamma}%
  \BibitemOpen
  \href@noop {} {}\bibinfo {note} {At present, we do not understand the reason
  why the electronic specific-heat coefficient $\gamma$ is not enhanced by the
  strong coupling effect in the tetragonal phase and is almost independent of
  phosphorus content $x$.}\BibitemShut {Stop}%
\bibitem [{\citenamefont {Carbotte}(1990)}]{RMP.62.1027}%
  \BibitemOpen
  \bibfield  {author} {\bibinfo {author} {\bibfnamefont {J.~P.}\ \bibnamefont
  {Carbotte}},\ }\href@noop {} {\bibfield  {journal} {\bibinfo  {journal} {Rev.
  Mod. Phys.}\ }\textbf {\bibinfo {volume} {62}},\ \bibinfo {pages} {1027}
  (\bibinfo {year} {1990})}\BibitemShut {NoStop}%
\bibitem [{\citenamefont {Hiroi}\ \emph {et~al.}(2007)\citenamefont {Hiroi},
  \citenamefont {Yonezawa}, \citenamefont {Nagao},\ and\ \citenamefont
  {Yamaura}}]{PhysRevB.76.014523}%
  \BibitemOpen
  \bibfield  {author} {\bibinfo {author} {\bibfnamefont {Z.}~\bibnamefont
  {Hiroi}}, \bibinfo {author} {\bibfnamefont {S.}~\bibnamefont {Yonezawa}},
  \bibinfo {author} {\bibfnamefont {Y.}~\bibnamefont {Nagao}}, \ and\ \bibinfo
  {author} {\bibfnamefont {J.}~\bibnamefont {Yamaura}},\ }\href@noop {}
  {\bibfield  {journal} {\bibinfo  {journal} {Phys. Rev. B}\ }\textbf {\bibinfo
  {volume} {76}},\ \bibinfo {pages} {014523} (\bibinfo {year}
  {2007})}\BibitemShut {NoStop}%
\bibitem [{\citenamefont {Nagao}\ \emph {et~al.}(2009)\citenamefont {Nagao},
  \citenamefont {Yamaura}, \citenamefont {Ogusu}, \citenamefont {Okamoto},\
  and\ \citenamefont {Hiroi}}]{JPSJ.78.064702}%
  \BibitemOpen
  \bibfield  {author} {\bibinfo {author} {\bibfnamefont {Y.}~\bibnamefont
  {Nagao}}, \bibinfo {author} {\bibfnamefont {J.}~\bibnamefont {Yamaura}},
  \bibinfo {author} {\bibfnamefont {H.}~\bibnamefont {Ogusu}}, \bibinfo
  {author} {\bibfnamefont {Y.}~\bibnamefont {Okamoto}}, \ and\ \bibinfo
  {author} {\bibfnamefont {Z.}~\bibnamefont {Hiroi}},\ }\href@noop {}
  {\bibfield  {journal} {\bibinfo  {journal} {J. Phys. Soc. Jpn.}\ }\textbf
  {\bibinfo {volume} {78}},\ \bibinfo {pages} {064702} (\bibinfo {year}
  {2009})}\BibitemShut {NoStop}%
\bibitem [{\citenamefont {Mine}\ \emph {et~al.}(2008)\citenamefont {Mine},
  \citenamefont {Yanagi}, \citenamefont {Kamiya}, \citenamefont {Kamihara},
  \citenamefont {Hirano},\ and\ \citenamefont {Hosono}}]{SSC.147.111}%
  \BibitemOpen
  \bibfield  {author} {\bibinfo {author} {\bibfnamefont {T.}~\bibnamefont
  {Mine}}, \bibinfo {author} {\bibfnamefont {H.}~\bibnamefont {Yanagi}},
  \bibinfo {author} {\bibfnamefont {T.}~\bibnamefont {Kamiya}}, \bibinfo
  {author} {\bibfnamefont {Y.}~\bibnamefont {Kamihara}}, \bibinfo {author}
  {\bibfnamefont {M.}~\bibnamefont {Hirano}}, \ and\ \bibinfo {author}
  {\bibfnamefont {H.}~\bibnamefont {Hosono}},\ }\href@noop {} {\bibfield
  {journal} {\bibinfo  {journal} {Solid State Commun.}\ }\textbf {\bibinfo
  {volume} {147}},\ \bibinfo {pages} {111} (\bibinfo {year}
  {2008})}\BibitemShut {NoStop}%
\end{thebibliography}
%

\end{document}